\documentclass[]{iopart}
\usepackage{graphicx}
\begin{document}

\title[Trapping photons by a line singularity]
 {Trapping photons by a line singularity}

\author{M. Arik   and O. Delice}
\address{ Bogazici University, Department of Physics, Bebek, Istanbul, Turkey}
\eads{\mailto{arikm@boun.edu.tr}, \mailto{odelice@boun.edu.tr}}
\begin{abstract}
We present cylindrically symmetric, static solutions of the
Einstein field equations around a line singularity such that the
energy momentum tensor corresponds to infinitely thin photonic
shells. Positivity of the energy density of the  thin shell and
the line singularity is discussed. It is also shown that thick
shells containing mostly radiation are possible by a numerical
solution.
\end{abstract}


\section{Introduction}
The cylindrically symmetric solutions of the Einstein field
equations have always been one of the more popular areas of
research in general relativity. The static and rotating cylinders,
collapsing and expanding cylindrical shells as well as their
static counterparts have been studied in the literature. Although
such objects are awaiting astrophysical discovery it is important
to investigate different types of objects since we do not  yet
know which type may be realized in the cosmos. Some examples of
these solutions are given in \cite{levicivita} - \cite{bicak}.
 A review of static solutions has been given in  \cite{bonnor}.
\cite{kramer} and \cite{kramerGonna} present solutions describing
two counter-propogating beams of light confined to a cylindrical
region. The energy momentum tensor is nonvanishing inside the
cylinder and the solution is smoothly matched to the outside
Levi-Civita metric. In this paper we will investigate solutions of
the Einstein equations giving static cylindrical thin shells
composed of massles particles around a line singularity. We will
also present an approximate solution giving static cylindrical
thick shell composed of equal amount of oppositely rotating
photons along  the angular dimension $\phi$ around a line
singularity.

\section{ Properties of the Cylindrically Symmetric Static Spacetime}\label{section2}

The general cylindrically symmetric static vacuum spacetime
satisfying Einstein's field equations is given by the Levi-Civita
metric \cite{levicivita}, which can be written in the form
\cite{herrera1}

\begin{equation}
ds^{2}=-\rho ^{4\sigma }dt^{2}+\rho ^{4\sigma (2\sigma -1)}(d\rho
^{2}+ P^2dp^{2})
+Q^2\rho ^{2(1-2\sigma )}dq^{2} \label{Lc}
\end{equation}
where $t$ and $\rho $ are the time and the radial coordinates with
the range $-\infty <t< \infty ,$ $0 \leq \rho < \infty$ and
$\sigma$, $P$ and $Q$ are real constants. The behaviour of the
metric components determine the nature of the coordinates $p$ and
$q$. Transforming the radius $\rho $ into a proper radius $r$ by
defining
\begin{equation}
dr=\rho ^{2\sigma (2\sigma -1)}d\rho
\end{equation}
puts the Levi-Civita metric (\ref{Lc}) in the Kasner form
\cite{herrera1}
\begin{equation}
ds^{2}=-R^{4\sigma
/N}dt^{2}+dr^{2}+P^2 R^{4\sigma (2\sigma -1)/N}dp^{2}
+Q^2 R^{2(1-2\sigma )/N}dq^{2} \label{Lc1}
\end{equation}
where

\begin{equation}\ \rho =R^{1/N},\ \  R=Nr, \ \ N=4\sigma ^{2}-2\sigma +1.
\end{equation}

One of the coordinates $p$ and $q$ has to be interpreted as the
axis of symmetry and the other as the angle about this axis. If we
fix the angular coordinate to a finite range then one of the
constants $P$ and $Q$ can be transformed away by a scale
transformation depending on the behaviour of the coordinates $p$
and $q$. This leaves the metric with only two independent
parameters. The parameter $\sigma$ is related to the energy of the
cylindrical source \cite{marder} whereas the surviving one of the
parameters $P$ and $Q$ is related to the topology of spacetime
\cite{Bonnorold}.

It is generally accepted that for small values of the parameter
$\sigma $ $\ (\sigma <\frac{1}{4})$, this metric corresponds to
cylindrically symmetric vacuum\ spacetime with $p$ as $z$
coordinate and $q$ as angular coordinate. Recently, this limit was
extended to the range $0\leq \sigma <\frac{1}{2}$ \cite{bonnor}.
Finally it was shown in \cite{herrera1} and \cite{herrera2} that
for $0\leq \sigma <\infty $ the Levi-Civita metric describes the
cylindrically symmetric \ vacuum spacetime provided that at
$\sigma =\frac{1}{2}$ the coordinates $p$ and $q$ interchange
their roles, that is, for $\sigma
>\frac{1}{2},$ $p$
becomes the angular coordinate and $q$ becomes the $z$ coordinate.
For $\sigma = \frac{1}{2},$ neither $p$ nor $q\ $ is entitled to
be an angular coordinate, and the three coordinates $(r,p,q)$ are
better visualized as Cartesian coordinates $x,y$ and $z$
\cite{herrera1},\cite{dasilva}. The $\sigma<0$ case is isotropic
to the $\sigma>0$ case.

In this paper we will thus use the cylindrically symmetric vacuum
metric in Kasner form. After rescaling the coordinates $t$ and $z$
and changing the constant in $g_{\phi \phi }$ to give correct
dimensions to the metric components we obtain

\begin{equation}
ds^{2}=-(\frac{r}{r_{0}})^{2a}dt^{2}+dr^{2}+(\frac{r}{r_{0}})^{2b}dz^{2}+(%
\frac{r}{r_{0}})^{2c}r_{0}^{2}\alpha ^{2}d\phi ^{2},
\label{kasner}
\end{equation}
where $\alpha ,a,b,c$ are real constants, $r$ is the radial
coordinate, $z$ is the axis of the line singularity and $\phi $ is
the angular coordinate with the range $0\leq\phi\leq 2\pi$. The
relation between parameters of the Levi-Civita form of the metric
and its equivalent Kasner form is given by
\begin{equation}
  a=\frac{2\sigma}{N},\ \ \
b=\frac{-2\sigma (1-2\sigma )}{N},\ \ 
c=\frac{1-2\sigma }{N}, \ \ \  N=4\sigma ^{2}-2\sigma +1.
\label{sigmawithabc}
\end{equation}
The dependence of $a,b$ and $c$ on $\sigma$ is shown in Figure 1.
\begin{figure}
\begin{center}
\includegraphics{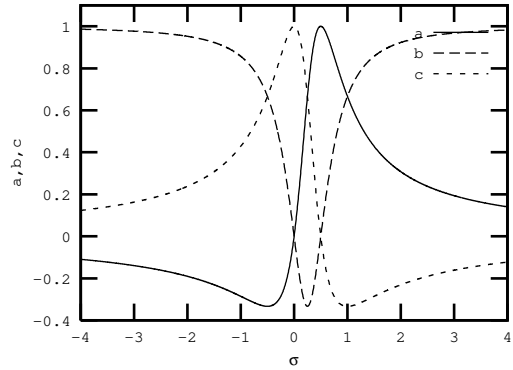}
\end{center}
\caption{\label{abcsigma}The dependence of $a,b,c$ on $\sigma$}
\end{figure}

 This metric is  cylindrically symmetric and
represents the exterior field of the line singularity at $r=0$
\cite{israel}. The Einstein tensor for this metric, apart from the
trivial solution ($a=b=c=0$), gives vacuum solutions of Einstein
equations with the constraints
\begin{equation}a+b+c=a^{2}+b^{2}+c^{2}=1,
\label{abcrel}
\end{equation}
so only one of the parameters $a,b,c$ is free. Choosing the free
parameter as $b$ than $a$ and $c$ are given by

\begin{equation}
a,c=\frac{(1-b)\pm \sqrt{(1-b)(1+3b)}}{2}
\end{equation}
where $\ \frac{-1}{3}\leq b\leq 1.$ For every value of $b$, $a$
and $c$ can take two  values. This is shown in Figure 2. In
general, the choice of $b$ as free parameter is arbitrary, so $
a,b,c$ can take values in between ($-\frac{1}{3}\leq a,b,c\leq 1$)
but only one of them can be negative at the same time. Note that,
the Riemann tensor vanishes only for the case where one of the
parameters is equal to one and the others are equal to zero.

\begin{figure}
\begin{center}
\includegraphics{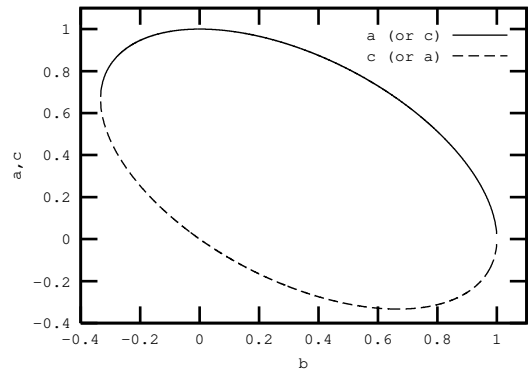}
\end{center}
\caption{\label{acwithb}The parameters $a$ and $c$ with $b$ as a
free parameter}
\end{figure}

The Levi-Civita metric (\ref{Lc1}) and its Kasner form
(\ref{kasner}) are equivalent. We thus discuss the relations
between $\sigma $ and $a,b,c$. When $0<\sigma <\frac{1}{2},$ \
$b<0,c>0$\ and when $\sigma
>\frac{1}{2},$ \ $b>0,c<0$. Since for $\sigma $ positive one
always has $a\geq 0,$ (Figure 1) one concludes that for the Kasner
form of the metric (\ref{kasner}),  for positive $\sigma$, the
coordinate corresponding to the parameter whose value is negative
has the role of $z$ coordinate and whose value is positive has the
role of angular coordinate of cylindrical coordinates. To avoid
confusion, for the rest of the paper, either we should consider
only $0\leq\sigma\leq 1/2$ which $b$ is always negative or we
should consider $0\leq \sigma<\infty$ which $b\ $ and $c\ $ change
sign at $\sigma=1/2$ (Figure 1). We choose latter and unless
stated explicitly, we write our expressions for $b<0$, but we
always have in mind that, for $b\ $ positive we have to replace
$b\ $ with $c\ $ in all metrics, expressions and figures without
writing them explicitly in our expressions. For negative $\sigma$,
same change of the roles of the coordinates may occur since for
small $\sigma, \ \ r^{2c}\approx r^2$ and for large $\sigma, \ \
r^{2b}\approx r^2$.

Note that the Levi-Civita metric in Kasner form (\ref{kasner}) has
another physically relevant parameter $\alpha$ whose value is
relevant to the topological defects in cylindrical coordinates.
For this metric, for $\sigma$ vanishing $a$ and $b$ vanish and $c$
becomes unity. On the other hand for $\sigma =\infty$, $a$ and $c$
vanish and $b$ becomes unity. In these cases our metric becomes:
\begin{equation} ds^{2}=-dt^{2}+dr^{2}+dz^{2}+\alpha^2 r^{2}d\phi
^{2}, \label{string}
\end{equation}
with $0\leq \phi \leq 2\pi $. This metric is flat and locally
Minkowskian but globally it has different topology and represents
a conical spacetime  for $\alpha \neq 1$ \cite{vilenkin}. For
$\alpha<1$ this metric gives the exterior field of an infinitely
long straight cosmic string lying along the $z$ direction.
Only $\alpha =1$ corresponds to flat and globally Minkowskian spacetime for $%
\sigma =0$ or $\sigma =\infty .$

We now turn to discuss the Newtonian Limit of the cylindrically
symmetric static Levi-Civita metric written in Kasner form
(\ref{kasner}). 
The Newtonian limit is given by $a=k\ll 1.$ Using the constraints
on $b$ $\&$ $c$ (\ref{abcrel})

\begin{equation}b+c=1-k
\end{equation}
\begin{equation}b^{2}+c^{2}=1-k
\end{equation}
there are two limits for $b$, $c$ in which one of them goes to
zero and the other to one. To identify the angular coordinate
correctly we have to choose $b\rightarrow 0$ and $c\rightarrow 1$.
Then
\begin{equation}
(r/r_0)^a=1+k \ln(r/r_0)+O(k^2)
\end{equation}
\begin{equation}
(r/r_0)^b=1-k\ln (r/r_0) + O(k^2)
\end{equation}
\begin{equation}
(r/r_0)^c=r/r_0+ O(k^2)
\end{equation}
Hence one obtains
\begin{equation}
ds^{2}=-(1+k\ln(r/r_{0}))^{2} dt^{2}+dr^{2}
+(1-k\ln (r/r_{0}))^{2}dz^{2}+\alpha^2 r^{2}d\phi ^{2}.
\end{equation}

Calculating the Einstein tensor for this metric verifies that it
is zero up to order $k^{2}.$ Thus one can identify the Newtonian
potential per unit mass as $V=k$ $ln(r/r_{0}).$ For positive $k $
this corresponds to an attractive force field while for
negative $k$ this corresponds to a repulsive force field    $%
\overrightarrow{g},$

\begin{equation}\overrightarrow{g}=-\overrightarrow{\nabla }V=-k\frac{\overrightarrow{r}}{r^{2}}. \label{force}
\end{equation}

A nonrelativistic particle in a circular orbit in this force field
has centripetal acceleration

\begin{equation}\frac{k}{r}=\frac{v^{2}}{r},
\end{equation}
so that the particle moves with constant velocity independent of
the radius. It is plausible that this generalizes to light which
always moves with constant velocity.

Note that the Levi-Civita metric in Kasner form ({\ref{kasner}) is
studied by Israel \cite{israel}. He showed that the singularity at
the $r=0$ represents the field of an infinite rod. He calculated
the line energy-momentum tensor density and showed that the source
has positive pressures and negative energy density. However the
effective gravitational mass per unit length is calculated as
$\frac{1}{2}a$ in our notation and is positive for positive $a\
(or \ \sigma)$ and negative for negative $a$. His results are in
accordance with the Newtonian limit of the metric (\ref{kasner})
where when $a$ is small and positive the field is attractive while
when $a$ is small and negative the force field is repulsive.

 For general relativity the metric (\ref{kasner}) admits helical null
 geodesics for a certain range of $\sigma$ with the angular velocity $\omega$ and the pitch velocity $v$ (the
constant velocity along $z$ direction of the  particles
 following a helical path). They are given by
\begin{equation}
\omega^2 = (\frac{a-b}{c-b})\ \frac{1}{\alpha^2 r_{0}^2}\
(\frac{r}{r_0})^{2(a-c)}, \label{geodesics}
\end{equation}
\begin{equation}
v^2=(\frac{c-a}{c-b})\ (\frac{r}{r_0})^{2(a-b)}.
\end{equation}
 For $b<0$ ($\sigma <1/2$), since $|c|\geq|b|$, there exist helical null geodesics only for
 $c\geq a$ ($0<\sigma\leq 1/4$) and for $a=c=2/3$ ($\sigma=1/4$) they become circular \cite{herrera1}. For
 $c<0$, $b>0$ ($\sigma >1/2$) helical null geodesics, which are circular for $a=b=2/3$ ($\sigma=1$), exist
 for $b\geq a$ ($\sigma\geq1$)(Figure 1).  Motivated by this, in
section 2 we will look for solutions of the Einstein equations
with an infinitely thin shell at $r=r_{1}.$ On the shell, the only
nonvanishing components of the energy momentum tensor satisfy
$T_\mu^\mu =0$ so that this shell can be interpreted as radiation
trapped in the gravitational field of the line singularity at
$r=0$.
The metric which is Ricci flat \ for $0<r<r_{1}$ and for $r>r_{1}$ is continuous at $%
r=r_{1}\ $ such that $T_{00}$ and $T_{\phi \phi },$ the only \
nonvanishing components of the energy-momentum tensor have Dirac
$\delta $-function singularities at $r=r_{1}.$ In section 4 we
will present an approximate thick shell solution  for the
$T_{00}=T_{\phi \phi }$ case such that the Einstein tensor is
finite and nonvanishing for $ r_1<r<r_2.$ We find that
$T_{00}=T_{\phi\phi}$ only approximately. The other diagonal
components of the energy momentum tensor also pick up small
values.

\section{Infinitely thin photonic shells around the line singularity}
\subsection{General case: A thin shell with counter moving photons along a helical path.}

There are several methods for calculating the Einstein tensor for
thin shell sources but the conditions  the metric should satisfy
are the same. The metric should be continuous everywhere but its
first derivatives   may be discontinuous  on  the shell and these
discontinuies give rise to an infinitely thin shell
\cite{israel2shell}-\cite{Taub}.

Using the same set of coordinates but different parameters for the
exterior and interior metrics we choose two different Levi-Civita
metrics in Kasner form (\ref{kasner}) for the interior and
exterior regions  of the infinitely long static thin shell with
radius $r_1$. These are given by :
\begin{equation}
ds_{-}^{2}=-(\frac{r}{r_{1}})^{2a'}dt^{2}+dr^{2}+(\frac{r}{r_{1}}
)^{2b'}dz^{2}
+(\frac{r}{r_{1}})^{2c'}\alpha^{2}r_{1}^{2} d\phi ^{2}\  \ \ \
(r<r_{1})\label{interiorkmetric}
\end{equation}
and
\begin{equation}
ds_{+}^{2}=-(\frac{r}{r_{1}})^{2a}dt^{2}+dr^{2}+(\frac{r}{r_{1}}
)^{2b}dz^{2}+(\frac{r}{r_{1}})^{2c}\alpha ^{2}r_{1}^{2} d\phi ^{2}
\ \ \ \ \ (r>r_{1}). \label{exteriorkmetric}
\end{equation}
where 
the constant parameters $a,b,c$ and $a',b',c'$ satisfy the
relations (\ref{abcrel}) so these metrics are both Ricci flat. As
explained in the introduction we must have $b$, $b'<0$ so that the
$z$ coordinate is identified correctly.

We can combine (\ref{interiorkmetric}) and (\ref{exteriorkmetric})
in the form :

\begin{equation}ds^{2}=-A^{2}(r)dt^{2}+dr^{2}+B^{2}(r)dz^{2}+D^{2}(r)d\phi
^{2} \label{metricABC}
\end{equation}
with
\begin{equation}A(r)=(r/r_{1})^{a'}\ \theta (r_{1}-r)+(r/r_{1})^{a}\ \theta (r-r_{1}),
\end{equation}

\begin{equation}B(r)=(r/r_{1})^{b'}\ \theta (r_{1}-r)+(r/r_{1})^{b}\ \theta (r-r_{1}), \end{equation}
and
\begin{equation}D(r)=[(r/r_{1})^{c'}\ \theta (r_{1}-r)+(r/r_{1})^{c}\ \theta (r-r_{1})]r_{1}\alpha ,
\end{equation}
where $\theta (x-x_{0})$ is the Heaviside step function with
\begin{eqnarray}\theta (x-x_{0})=0, \ \ \ \ x<x_{0} \nonumber \\
\theta (x-x_{0})=1,\ \ \ \ x\geq x_{0}
 \end{eqnarray}

At $r=r_{1}$ this metric is continuous but its first derivatives
with respect to $r\ $ are discontinuous. The nonzero components of
the Einstein tensor for the metric (\ref{metricABC}) are given by

\begin{equation}G_{00}=-(\frac{B_{rr}}{B}+\frac{D_{rr}}{D}+\frac{B_{r}D_{r}}{BD})
\end{equation}

\begin{equation}G_{rr}=G_{11}=\frac{A_{r}B_{r}}{AB}+\frac{A_{r}D_{r}}{AD}+\frac{B_{r}D_{r}}{
BD}
\end{equation}

\begin{equation}G_{zz}=G_{22}=\frac{A_{rr}}{A}+\frac{D_{rr}}{D}+\frac{A_{r}D_{r}}{AD}
\end{equation}

\begin{equation}G_{\phi \phi }=G_{33}=\frac{A_{rr}}{A}+\frac{B_{rr}}{B}+\frac{A_{r}B_{r}}{AB
},
\end{equation}
where subscripts denote the partial derivatives.
%
Since the interior and the exterior regions of the shell is
vacuum, the only surviving terms are the terms which contain Dirac
delta functions which give the energy momentum tensor of the
shell.

So for $b$, $(b')<0$ case the nonzero elements of $G_{\mu \nu }$
are

\begin{equation}G_{00}=-\frac{b-b'+c-c'}{r_{1}}\ \delta (r-r_{1})=\frac{%
a-a'}{r_{1}}\ \delta (r-r_{1}),
\end{equation}

\begin{equation}G_{22}=\frac{a-a'+c-c'}{r_{1}}\ \delta (r-r_{1})=\frac{%
b'-b}{r_{1}}\ \delta (r-r_{1}), \label{G22shell}
\end{equation}

\begin{equation}G_{33}=\frac{a-a'+b-b'}{r_{1}}\ \delta (r-r_{1})=\frac{%
c'-c}{r_{1}}\ \delta (r-r_{1})\label{G33shell}.
\end{equation}

For the Einstein equation
\begin{equation}
G_{\mu\nu}=8 \pi G \ T_{\mu\nu}
\end{equation}
we can choose the energy momentum tensor of the shell in the form
\begin{equation}
T_{\mu\nu}=diag(\rho,p_r,p_z,p_\phi)\label{enmomtensor}
\end{equation}
where $\rho$ is the energy density  and $p_i \ (i=1,2,3)$ are the
principal pressures \cite{HE}. Since we have $T^0_0=-T_{00}$ and
$T^i_j=T_{ij}\ (i,j=r,z,\phi) $, using (\ref{abcrel}) one can show
that the energy momentum tensor of the shell satisfies the
condition $T^\mu_\mu =0$ and this result can be interpreted as an
infinitely long thin shell along the $z$ direction with radius
$r_1$ composed of equal amount of oppositely moving photons along
a helical direction. This helical motion gives rise to pressures
in the $\phi$ and $z$ directions with the equation of state
$\rho=p_z+p_\phi.$ Thus if one chooses the interior and the
exterior metrics of the shell as Levi-Civita metrics in Kasner
form (\ref{interiorkmetric},\ref{exteriorkmetric}) then the shell
is necessarily composed of massles particles.

 Energy per unit length of the shell is given by :
 \begin{equation} \mu= \alpha(a-a') .\end{equation}

For $(a-a')>0$ the shell has positive energy density. 
 Since the interior metric is of Kasner form, we have a line
singularity at $r=0$. The singularity has positive effective mass
density for $a'$ positive and negative effective mass density for
negative $a'$. The only singularity free metric is flat Minkowski
metric and if we take interior metric (\ref{interiorkmetric}) as
Minkowski metric (\ref{string}) ($a' =b' =0,c' =1, \alpha =1$ )
then our results
are in accordance with 
\cite{herrera1} and \cite{bicak}.
%
In this case the shell has positive energy density for $a$
positive and satisfies all energy conditions for $|a|>|b|.$

\subsection{A thin shell with counter-rotating photons}

Now we would like to discuss the case which may be called the
photon cylinder with counter-rotating photons. We find a static
cylindrical shell at $r=r_1$ with $T_{00}= \rho =p_{\phi }=T_{22}$
with other components of the energy momentum tensor vanishing. For
comparison we note that for light in a laser tube and for two
counter-propagating beams of light
\cite{kramer}-\cite{kramerGonna} the equation of state is
$\rho=p_z$. We take $G_{22}=0$ in (\ref{G22shell}), which gives:


\begin{equation}b=b'.
\end{equation}

Taking b as the free parameter , the parameters $a, c, a'$ and
$c'$ are given by
\begin{equation}
a,c=a' ,c'=\chi _{+,}\chi _{-} =\frac{(1-b)\pm \sqrt{(1-b)(1+3b)}
}{2}.
\end{equation}
For this case to have positive energy density on the shell we need
$a-a'>0$.
Since $a>a'$ leads to $c'>c$ then we have to choose $a' =c=\chi
_{-}$ and $ c'=a= \chi _{+}$. Otherwise the shell dissappears
($G_{00}=G_{33}=0$).

Applying these relations to (\ref{interiorkmetric}) and
(\ref{exteriorkmetric}) our interior and exterior metrics become
\begin{equation}
ds_{-}^{2}=-(\frac{r}{r_{1}})^{2c}dt^{2}+dr^{2}
+(\frac{r}{r_{1}})^{2b}dz^{2}+ (\frac{r}{r_{1}})^{2a}\alpha ^{2}\
r_{1}^{2}d\phi ^{2}\ \ \  (r<r_{1})\label{intrmetrshel}
\end{equation}
and
\begin{equation}
ds_{+}^{2}=-(\frac{r}{r_{1}})^{2a}dt^{2}+dr^{2} 
+(\frac{r}{r_{1}})^{2b}dz^{2} +(\frac{r}{r_{1}})^{2c}\alpha ^{2}\
r_{1}^{2}d\phi ^{2} \ \ \  (r>r_{1})\label{extrmetrshel}
\end{equation}
and the only nonzero elements of the Einstein tensor are
\begin{equation}
G_{00}=G_{\phi \phi }=\frac{c^{\prime }-c}{r_{1}}\ \delta
(r-r_{1})=\frac{
a-a^{\prime }}{r_{1}}\ \delta (r-r_{1}) 
=\frac{a-c}{r_{1}}\ \delta (r-r_{1}).\label{enofphotshel}
\end{equation}
\begin{figure}
\begin{center}
\includegraphics{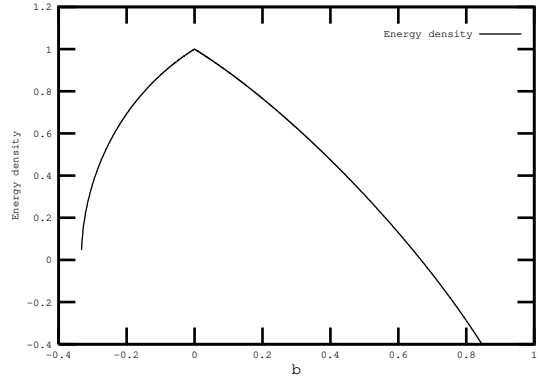}
\end{center}
\caption{\label{eos} Changing of the energy density
(\ref{enofphotshel}) of the shell with counter-rotating photons
with $b$.}
\end{figure}
It is clear that a shell with positive energy requires the
condition $a>c$ for $c>0$. Then our shell has positive energy
density only for $\frac{1}{4}<\sigma <1$ (see figure(1))  or
equivalently $-1/3<b<2/3$ and $a>c$ (see figure 2) for the
external metric (\ref{exteriorkmetric}). For interior metric
(\ref{interiorkmetric}) these conditions correspond to $0<\sigma
'<1/4$ or $\sigma ' >1$ since $c'=a$ and $a'=c$. For these values
of $\sigma$, one of the quantities $b$ and $c$ in (\ref{abcsigma})
is negative. In accordance with the interpretation of the $z$
coordinate we choose name the negative one as $b\ $ in expressions
above (For $b>0$ case see  Section \ref{section2}
). The energy
density of this cylindrical shell has a finite maximum value at
$b=0$ (Figure 3). This is consistent with other static cylindrical
shell solutions \cite{herrera1}. Note that the interior metric
(\ref{intrmetrshel}) always admits helical null geodesics but
exterior one (\ref{extrmetrshel}) does not.

The energy per unit length of this shell is:
\begin{equation}
\mu =\alpha (a-c).
\end{equation}

As we mentioned before, unless we choose globally Minkowski metric
inside the shell , at $r=0 $ there is a line singularity. When the
shell satisfies positive energy condition the singularity at the
origin has positive effective mass density. The singularity free
configuration for the interior metric is given by $c=b=0$, $a=1$,
$\alpha=1$ (\ref{string}) which describes a cylindrical shell with
flat interior.
 The exterior metric becomes

\begin{equation}ds^{2}=-(\frac{r}{r_{1}})^{2}dt^{2}+dr^{2}+dz^{2}+r_{1}^{2}d\phi
^{2}. \label{rindler}
\end{equation}

This metric is the Rindler metric \cite{dasilva, HerRuiSan}, which
represents a uniform gravitational field and it is Riemann flat.
The test particles are uniformly accelerated in this field. The
gravitational field of a massive plane in Newtonian theory is also
uniform. This extreme case happens when energy density of the
cylinder becomes maximum (see Figure (\ref{eos})). One can
conclude that for this case the cylinder becomes an infinite
plane.

\subsection{A thin shell with counter moving photons in the $z-$direction}
 In this case we find a shell with the equation of state $\rho=p_z$. we choose $c=c'$ in (\ref{G33shell}) which leads to
 $a=b'$and $a'=b$ then the interior and exterior metrics  become:
\begin{equation}
ds_{-}^{2}=-(\frac{r}{r_{1}})^{2b}dt^{2}+dr^{2}
+(\frac{r}{r_{1}})^{2a}dz^{2}+ (\frac{r}{r_{1}})^{2c}\alpha ^{2}\
r_{1}^{2}d\phi ^{2}\ \ \  (r<r_{1})\label{intrmetrzshel}
\end{equation}
and
\begin{equation}
ds_{+}^{2}=-(\frac{r}{r_{1}})^{2a}dt^{2}+dr^{2} 
+(\frac{r}{r_{1}})^{2b}dz^{2} +(\frac{r}{r_{1}})^{2c}\alpha ^{2}\
r_{1}^{2}d\phi ^{2} \ \ \  (r>r_{1})\label{extrmetrzshel}
\end{equation}
and the only nonzero elements of the Einstein tensor are
\begin{equation}
G_{00}=G_{zz}=\frac{b^{\prime }-b}{r_{1}}\ \delta (r-r_{1})=\frac{
a-a^{\prime }}{r_{1}}\ \delta (r-r_{1}) 
=\frac{a-b}{r_{1}}\ \delta (r-r_{1}).\label{enofzphotshel}
\end{equation}

To have a positive energy density of the shell we must have
$a-b>0.$ Since when $a>0$, $b(=a')<0$, the shell satisfies all
energy conditions. 
 Note that for this solution, unlike the first two cases, the
singularity at $r=0$ has negative effective mass density according
to \cite{israel} when the shell satisfies the positive energy
conditions. The energy density is finite (Figure \ref{eozs}). The
exterior metric (\ref{extrmetrzshel}) of the physically acceptable
shell may admit helical null geodesics but interior metric
(\ref{intrmetrzshel}) does not. The energy per unit lenght of the
shell is $\mu=\alpha (a-b).$ If the interior is chosen as Riemann
flat ($a=b=0, c=1$) then this shell disappears. 
\begin{figure}
\begin{center}
\includegraphics{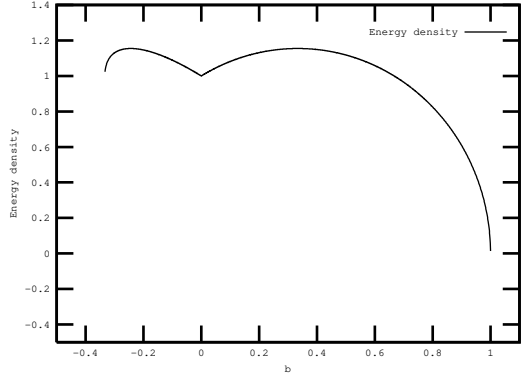}
\end{center}
\caption{\label{eozs} Changing of the energy density
(\ref{enofzphotshel}) of the shell with counter-moving photons
with $b$.}
\end{figure}


\section{Thick cylindrical shell around the line singularity: An approximate solution}

We try to replace the infinitely thin shell solution composed of
rotating photons  around a line singularity with a solution with a
smooth hollow cylindrical material with inner radius $r_{1}$ and
outer radius $r_{2}$. around a line singularity. We again choose
the metrics of the inside and the outside of the cylinder to be
the same as that of our thin shell solution
(\ref{interiorkmetric}), (\ref{exteriorkmetric}) except rescaling
the coordinates and adding some constants to satisfy the
continuity. In between we use a third metric which may represent
the cylindrical material.
\begin{equation}
ds_{-}^{2}=-r^{2c}dt^{2}+dr^{2}+r^{2b}dz^{2}
+\alpha^2 r^{2a}d\phi ^{2}  \ \ \ (r<r_1),
\end{equation}
\begin{equation}
ds_{0}^{2}=-e^{2}(r)dt^{2}+dr^{2}+f^{2}(r)dz^{2}
+\alpha^2 g(r)^{2}d\phi ^{2}  \  (r_1\leq r \leq r_2),
\label{metrefg}
\end{equation}
\begin{equation}
 ds_{+}^{2}=-C^2 r^{2a}dt^{2}+dr^{2}+ r^{2b}dz^{2}
 +\alpha^2 F^2 r^{2c}d\phi ^{2}
\ \ (r_2<r).
\end{equation}

 We  thus want the cylindrical material to be smooth
so the metric and its first derivatives should be continuous at
$r=r_{1}$ and $r=r_{2}.$ Then we find following boundary
conditions:

\begin{eqnarray}
&&\begin{tabular}{|l|l|l|l|} \hline & $e(r)$ & $f(r)$ & $g(r)$
\\ \hline $r=r_{1}$ & $r_{1}^{c}$ & $r_{1}^{b}$ & $r_{1}^{a}$ \\
\hline $r=r_{2}$ & $Cr_{2}^{a}$ & $r_{2}^{b}$ & $Fr_{2}^{c}$ \\
\hline
\end{tabular} \ \ \ \ \nonumber \\
&&\begin{tabular}{|l|l|l|l|} \hline & $e_r(r)$ & $ f_r(r)$ &
$g_r(r)$ \\ \hline $r=r_{1}$ & $cr_{1}^{c-1}$ & $br_{1}^{b-1}$ &
$ar_{1}^{a-1}$ \\ \hline $r=r_{2}$ & $Car_{2}^{a-1}$ &
$br_{2}^{b-1}$ & $Fcr_{2}^{c-1}$ \\ \hline
\end{tabular}%
\ \ \ \
\end{eqnarray}

where $C$ and $F$ are constants to be determined from these
boundary conditions. Since we have to match the metric and its
first derivative at the boundaries we simply take $e(r)$ and
$g(r)$ to be quadratic functions whose coefficients are determined
by the boundary conditions. Since the $dz^2$ part of the metric is
the same inside and outside we also choose $f=r^b$. We use the
following ansatz for $e(r),f(r)$, and $g(r)$\ and we replace
$r_{1}=m$ and $r_{2}=n$:
\begin{equation}
 e(r)=cm^{c-1}r+A+\frac{(r-m)^{2}}{(n-m)^{2}}
 (Can^{a-1}-(cm^{c-1}n+A)+B),
\end{equation}
\begin{equation}f(r)=r^{b} ,\ 
\end{equation}
\begin{equation}g(r)=Fcn^{c-1}r+E+\frac{(r-n)^{2}}{(m-n)^2}G,
\end{equation}
where

\begin{equation}
A=m^c(1-c),
\end{equation}
\begin{equation}
B=Cn^{a-1}(n-a),
\end{equation}
\begin{equation}
C=\frac{cm^{c-1}(n-m)+2m^c}{an^{a-1}(m-n)+2n^a},
\end{equation}

\begin{equation}
E=Fn^c(1-c),
\end{equation}
\begin{equation}
G=-F(cn^{c-1}(m-n)+n^c)-m^a,
\end{equation}
and
\begin{equation}
F=-n\frac{2m^a+am^{a-1}(n-m)}{cn^{c-1}(n-m)-2n^2}.
\end{equation}

Next we calculate  the nonzero elements of the Einstein tensor
between $m<r<n$ for $m=1$ and $n=2$.
 We refer the reader to the appendix for the explicit expressions. Here we
present a series of graphs showing the components of the Einstein
tensor as a function of $r$ for seven different values of $b$
(Figure 4-10). We consider the energy momentum tensor of this
thick shell in the form (\ref{enmomtensor}). Note that for the
thin shell solution we have only $\rho =p_{\phi }\neq 0.$ For the
thick shell case, since we present an approximate solution, the
pressure in other directions also contribute to the
energy-momentum tensor.
 We show that their contribution is small compared to the energy
density and azimuthal stress. The radial pressure $p_{r}$ should
vanish at the boundaries.

Note that \ for $-1/3<b\leq -0.235$ and $0.315\leq b\leq 1$ our
solution
satisfies only the weak energy condition since at some region $p_{\phi }>\rho $ . For $%
-0.235<b<0.315$ all energy conditions are satisfied. \nopagebreak
As $b$ goes from -$1/3$ to $0$\ the energy density increases and
reaches a finite maximum value at $b=0$. Then as $b$ increases to
$1$ the energy decreases towards zero. Having a maximum finite
energy density with $b$
 is an expected behaviour since previous shell solutions \cite{bonnor} have this behaviour too.
 If we accept the dominant
energy condition as physically relevant energy condition, because
otherwise the local speed of sound can be greater than the speed
of light \cite{HE} when $p_{i}>\rho ,$ then our solution is
physically acceptable only for $-0.235<b<0.315.$ \ \ \ \ \
\bigskip
\begin{figure}
\begin{center}
\includegraphics{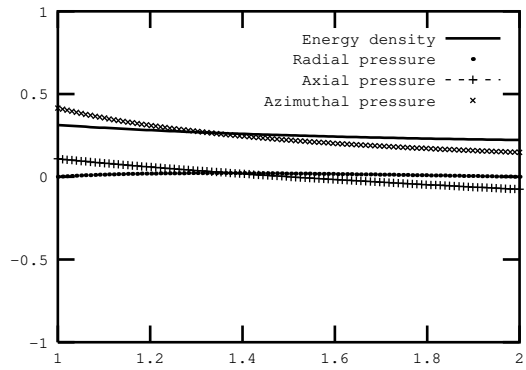}
\end{center}
\caption{\label{b-3}The nonzero components of the Einstein tensor
for $b=-0.3$ of the metric (\ref{metrefg})}
\end{figure}

\begin{figure}
\begin{center}
\includegraphics{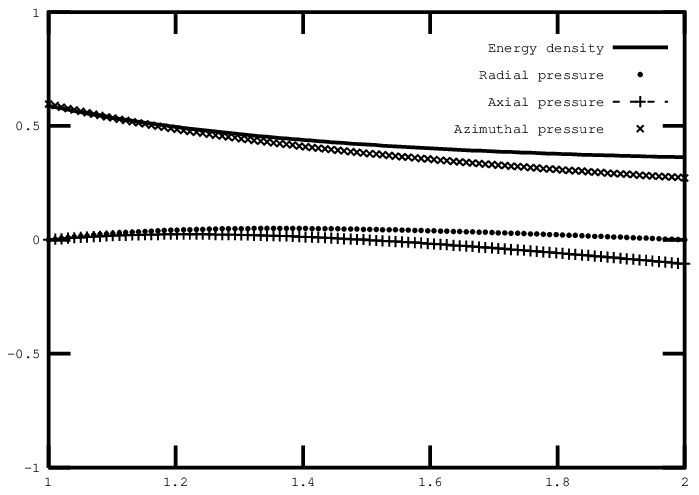}
\end{center}
\caption{\label{b-235}The nonzero components of the Einstein
tensor for $b=-0.23$ of the metric (\ref{metrefg})}
\end{figure}


\begin{figure}
\begin{center}
\includegraphics{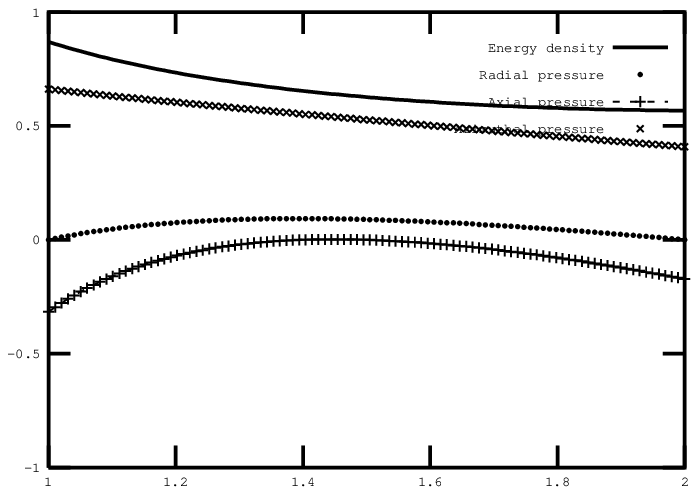}
\end{center}
\caption{\label{b-1}The nonzero components of the Einstein tensor
for $b=-0.1$ of the metric (\ref{metrefg})}
\end{figure}

\begin{figure}
\begin{center}
\includegraphics{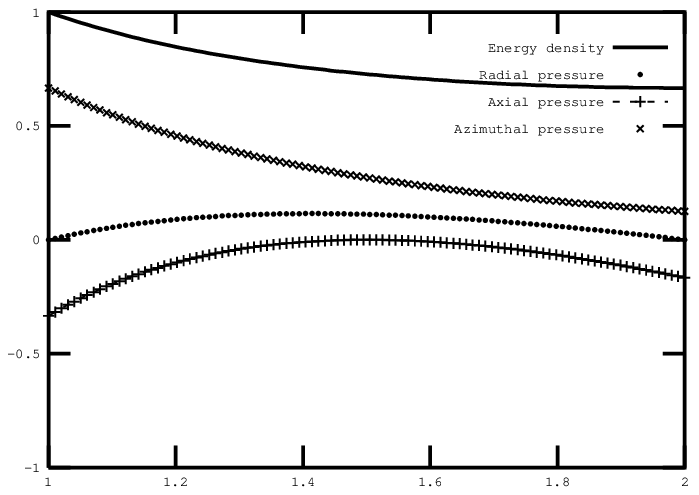}
\end{center}
\caption{\label{b0}The nonzero components of the Einstein tensor
for $b=0$ of the metric (\ref{metrefg})}
\end{figure}


\begin{figure}
\begin{center}
\includegraphics{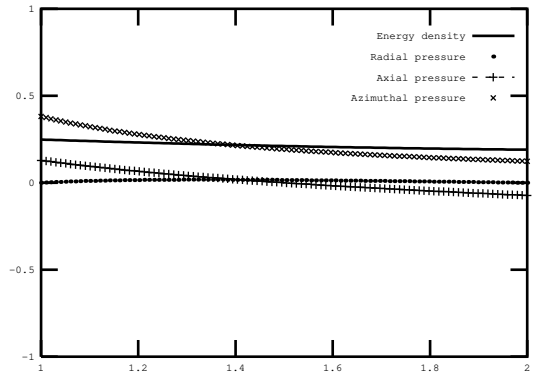}
\end{center}
\caption{\label{b05}The nonzero components of the Einstein tensor
for $b=0.5$ of the metric (\ref{metrefg})}
\end{figure}

\section{CONCLUSION}
\bigskip
In this paper, we investigated static, cylindrically symmetric
solutions of Einstein equations for an infinitely thin cylindrical
shell around a line singularity with $T^\mu_\mu=0\ $ on the shell.
We presented three cases which correspond to counter-propagating
photons along a helical direction, along the circular direction
and parallel to the cylinder axis. For the third case we found
that the line singularity cannot have positive mass density. The
interior and exterior regions of this cylindrically symmetric
infinitely long shell are two different Levi-Civita vacuum
spacetimes of Kasner form. The existence of circular and helical
null geodesics for this metric reinforces our interpretation. Then
we presented an approximate solution for the thick shell case with
counter rotating photons  around a line singularity where the
other diagonal stresses also contribute a little to the
energy-momentum tensor.

 We have choosen the metric interior to the shell  in Kasner
form (\ref{kasner}) which has a line singularity at $r=0$
\cite{israel} and have shown that the Newtonian limit of this
metric corresponds to an attractive force field (\ref{force}) for
positive $a$ and repulsive force field for negative $a$. This
result is consistent with Israel's analysis of the same metric
\cite{israel}.


It has been demonstrated in \cite{amster},\cite{peter} and
\cite{ray} that the singularity at the $r=0$ of the cylindrically
symmetric, static metric (\ref{kasner}) can represent a
superconducting cosmic string. The outside of the infinitely long
straight ordinary cosmic string along the $z$ direction can be
represented by a locally flat Minkowski metric with an angular
deficit \cite{vilenkin}. In this case there is a boost invariance
along the $z$ direction and this either corresponds to $a=b=0$ and
$c=1$ or to the case $a=b=\frac{2}{3},c=\frac{-1}{3 }$ which is
also labeled as the external field of a static string \cite{ray},
\cite{vilenkin}. However, considering \cite{herrera1} and
\cite{herrera2} one realizes that $c$ cannot be negative if it
represents the parameter related to $g_{\phi \phi }$ of the metric
(\ref{kasner}). Thus we exclude this solution from our
considerations. For the superconducting straight string case it
was shown that the effects due to current are negligible
\cite{peter} and the external field can be represented by a metric
in Kasner form  \cite{amster} at least asymptotically. In this
case the boost invariance along the $z$ direction is broken by the
currents and in general we can take $a\neq b.$ The deficit angle
is not sensitive to the value of current but decreases when the
string is about to switch from the superconducting phase to the
ordinary one. Considering these arguments, we can say that the
singularity at the $r=0$ is consistent with a superconducting
cosmic string. For the ordinary cosmic string the Rienmann tensor
vanishes but for the superconducting case it does not. The
ordinary cosmic string does not have helical null geodesics but
the superconducting one has for $\sigma\leq 1/4$ or $1\leq\sigma$.
For $\sigma=1/4$ or $1$ they become circular.

\cite{Dyer} investigated a line singularity such that the
energy-momentum tensor is given by a Nielsen-Olesen \cite{nielsen}
vortex. They concluded that there is no photon cylinder around
their string solution which in our solution corresponds to the
$a=b$ case where we find that there are no thin or thick shell
solutions satisfying the energy conditions.

 For $-1/3<b<2/3$ the thin shell solution with the equation of state
$\rho=p_\phi$ is physically relevant since it obeys all energy
conditions for this range of $b$.  For the exterior metric
(\ref{exteriorkmetric}) this range is equal to $1/4 < \sigma < 1$.
For the thick shell solution the dominant energy condition is
satisfied only for $-0.235<b<0.315$ and for this range of $b$ this
solution can be  physically acceptable. As we change the parameter
b form  $-1/3$ to 1 the energy density of this thin shell solution
first increases then reaches a finite maximum at $b=0$ and then
decreases (Figure \ref{eos}). The energy density of the other two
solutions are also finite.

  This is the expected behaviour according to the Hoop conjecture since the cylindrical
 collapse does not produce a horizon due to the fact that the source is not
 bounded from every direction. 
It is also interesting that for the maximal energy density case
$b=0$, the "cylinder" surrounding the line singularity becomes a
flat plane when observed from the outside i.e. the cylindrical
shell becomes a cosmic wall.

\section*{Appendix}
\appendix
\setcounter{section}{1}
Here we give the expressions of the nonzero  components of the
Einstein tensor of the metric (\ref{metrefg}) for $r_1=1$, $r_2=2$
for different  values of the parameters $a,b,c$

$b=-0.3$, $a=0.83$, $c=0.47$  (Figure 4) \
\begin{eqnarray} \nonumber  G_{00}&=&\frac{-1. 8r^2+0.52r-0.095}{r^2(r-0.044)(r-5.6)}\\
 \nonumber   G_{rr}&=&\frac{ 2.8\,(r+2.1)\,(r-1)(r-2)}{r\, (r-0.045)\,(r-5.6)\,(r^2+3.3\,r+7)} \\
\nonumber  G_{zz}&=&\frac{-4.2-9.2\ r+8\ r^2}{(r^2+3.3r+7.0)(r-0.044)(r-5.6)} \\
  G_{\phi\phi}&=&\frac{2.6+0.29\ r+1.8\ r^2}{r^2(r^2+3.3\ r+7.0)}
\end{eqnarray}

$b= -0.23$, $a= 0.92$, $c=0.32$ (Figure 5)
\begin{eqnarray}
 \nonumber G_{00}&=&\frac{-1.8\, r^2+0.26\, r-0.22}{r^2\,(r-0.17)(r-4.6)}\\
\nonumber  G_{rr}&=&\frac{3.1\, (r+0.85)\, (r-1)\, (r-2)}{r\, (r-0.17)\, (r-4.6)\, (r^2-0.32\, r+4.6)}\\
\nonumber  G_{zz}&=&\frac{8\, (r-1)\,(r-1.5)}{(r-0.17)\,(r-4.6)\,(r^2-0.32\,r+4.6)}\\
 G_{\phi\phi}&=&\frac{ 25-0.34\  r+35\ r^2}{(87-6 \ r+19\ r^2)r^2}
\end{eqnarray}


 $b=-0.1$, $a=0.99$, $c=0.11$  (Figure 7)
\begin{eqnarray}
\nonumber    G_{00} &=& \frac{-22+8.7\ r-430\ r^2}{r^2\ (200-940\ r+230\ r^2)} \\
\nonumber    G_{rr}&=&\frac{ 3.6\, (r+0.26)(r-1)(r-2)}{r\, (r-0.25)\, (r-3.9)(r^2-1.6\, r+4.1)}  \\
\nonumber  G_{zz}&=& \frac{8\, (r-1.4)\, (r-1.5)}{(r-0.25)\, (r-3.9)(r^2-1.6\, r+4.1)} \\
 G_{\phi\phi}&=&\frac{13-0.47\ r+55\ r^2}{r^2\ (120-47\ r+29\ r^2)}
\end{eqnarray}

$b=0$, $a=1$, $c=0$   (Figure 8)
\begin{eqnarray}
\nonumber G_{00}&=&\frac{-2}{1-4\ r+r^2}\\
\nonumber G_{rr}&=&\frac{ 4(r-1)(r-2)}{(4-2r+r^2)(1-4r+r^2)}\\
\nonumber G_{zz}&=&\frac{18-24r+8r^2}{(4-2r+r^2)(1-4r+r^2)}\\
 G_{\phi\phi}&=& \frac{2}{4+r^2-2r}
\end{eqnarray}


 $b=0.5$, $a=0.81$, $c=-0.31$    (Figure 10)
\begin{eqnarray}
 \nonumber G_{00}&=&\frac{-5.2+120\ r-360\ r^2}{r^2\ (13-1200\ r+200\ r^2)}\\
 \nonumber G_{rr}&=&\frac{2.8\ (r+2.8)\ (r-1)(r-2)}{r\ (r-0.011)\ (r-6)\ (r^2+5.5r+8.5)}\\
 \nonumber G_{zz}&=&\frac{8.2\ (r+1.3)\ (r-1.5)}{(r-0.022)\ (r-6)\ (r^2+5.5\ r+8.5)}\\
 G_{\phi \phi}&=&\frac{1.8\ r^2+0.52\, r+3.4}{r^2\ (r^2+5.5\, r+8.5)}
\end{eqnarray}

\section*{Acknowlegment}
 We would like to thank C. Saclioglu
for discussions.
\bibliographystyle{amsplain}
\bibliography{xbib}


\end{document}